\documentclass[conference]{IEEEtran}
\IEEEoverridecommandlockouts
\usepackage{cite}
\usepackage{amsmath,amssymb,amsfonts}
\usepackage{algorithmic}
\usepackage{graphicx}
\usepackage{textcomp}

\usepackage{lipsum}
\usepackage{comment}
\usepackage{multirow}
\usepackage{subfigure}
\usepackage{bm}
\usepackage{xcolor}
\usepackage{url}
\usepackage{xcolor}
\usepackage{tikz}
\usepackage{textcomp}
\usepackage{hyperref}
\usepackage{fancyhdr}

\fancypagestyle{allpagesstyle}
{
    
    \lhead{\footnotesize THIS PAPER HAS BEEN ACCEPTED AT THE \underline{IEEE GCAIoT 2025} CONFERENCE AND FINAL VERSION IS SUBMITTED.}
    \rhead{\footnotesize \thepage}
    \cfoot{}
}
\newcommand\copyrighttext{
    \footnotesize \textcopyright 2025 IEEE. The final published paper is copyrighted by IEEE and should be cited as: Cade Kennedy, Amr Hilal, Morteza Momeni, ``The Role of Federated Learning in Improving Financial Security: A Survey,'' \textit{IEEE Global Conference on Artificial Intelligence \& Internet of Things (GCAIoT)}, 2025, pp. 1–8. Personal use of this material may be permitted and permission from IEEE must be obtained for all other uses.
}
\newcommand\copyrightnotice{
    \begin{tikzpicture}[remember picture,overlay]
    \node[anchor=south,yshift=10pt] at (current page.south){\fbox{\parbox{\dimexpr\textwidth-\fboxsep-\fboxrule\relax}{\copyrighttext}}};
    \end{tikzpicture}%
}

\def\BibTeX{{\rm B\kern-.05em{\sc i\kern-.025em b}\kern-.08em
    T\kern-.1667em\lower.7ex\hbox{E}\kern-.125emX}}
    
\begin{document}

\title{The Role of Federated Learning in Improving Financial Security: A Survey\\
}

\author{\IEEEauthorblockN{Cade Houston Kennedy}
\IEEEauthorblockA{\textit{Department of Computer Science} \\
\textit{Tennessee Technological University}\\
Cookeville, USA \\
chkennedy42@tntech.edu}
\and
\IEEEauthorblockN{Amr Hilal}
\IEEEauthorblockA{\textit{Department of Computer Science} \\
\textit{Tennessee Technological University}\\
Cookeville, USA \\
ahilal@tntech.edu}
\and
\IEEEauthorblockN{Morteza Momeni}
\IEEEauthorblockA{\textit{Department of Finance} \\
\textit{Tennessee Technological University}\\
Cookeville, USA \\
mmomeni@tntech.edu}
}

\maketitle
\copyrightnotice
\thispagestyle{allpagesstyle}

\begin{abstract}
With the growth of digital financial systems, robust security and privacy have become a concern for financial institutions. Even though traditional machine learning models have shown to be effective in fraud detections, they often compromise user data by requiring centralized access to sensitive information. In IoT-enabled financial endpoints such as ATMs and POS Systems that regularly produce sensitive data that is sent over the network. Federated Learning (FL) offers a privacy-preserving, decentralized model training across institutions without sharing raw data. FL enables cross-silo collaboration among banks while also using cross-device learning on IoT endpoints. This survey explores the role of FL in enhancing financial security and introduces a novel classification of its applications based on regulatory and compliance exposure levels— ranging from low-exposure tasks such as collaborative portfolio optimization \cite{b16} to high-exposure tasks like real-time fraud detection \cite{b7,b8}. Unlike prior surveys, this work reviews FL’s practical use within financial systems, discussing its regulatory compliance and recent successes in fraud prevention and blockchain-integrated frameworks. However, FL's deployment in finance is not without challenges. Data heterogeneity, adversarial attacks, and regulatory compliance make implementation far from easy. This survey reviews current defense mechanisms and discusses future directions, including blockchain integration, differential privacy, secure multi-party computation, and quantum-secure frameworks. Ultimately, this work aims to be a resource for researchers exploring FL’s potential to advance secure, privacy-compliant financial systems.
\end{abstract}

\section{Introduction}
Over the past decade, Machine Learning is a technology that has been utilized in several different industries, such as healthcare, transportation, and manufacturing. The finance industry has recently adopted this technology for several different use cases, most prominently, fraud detection \cite{b29,b30,b31}, loan approvals \cite{b32}, and stock market analysis \cite{b33}. Unfortunately, the evolution of this technology is heavily reliant on rich and high-quality datasets to train a highly accurate machine learning model \cite{b1}. This need for data has shifted the focus to obtaining data for these models. However, due to the highly sensitive nature of financial data, collecting large datasets often requires collaboration between financial institutions. This inter-institutional data sharing introduces significant privacy and security challenges regarding transparency of data collection, storage, and usage \cite{b2}. 

Federated Learning (FL) is a machine learning framework that opens the avenue of decentralized training on user-specific data where client data never leaves the device or institution in this case. In particular, cross-silo federated learning refers to the collaboration between a small number of trusted institutions (such as banks or financial firms) that jointly train a shared model without exposing their local datasets. In contrast, cross-device FL applies FL at the edge level like mobile banking apps, or POS terminals allowing financial institutions to make real-time predictions without pulling sensitive data from individual endpoints. This is particularly important as IoT-enabled financial devices generate large volumes of highly sensitive data. 

\begin{figure}[htbp]
    \centering
    \includegraphics[width=\columnwidth]{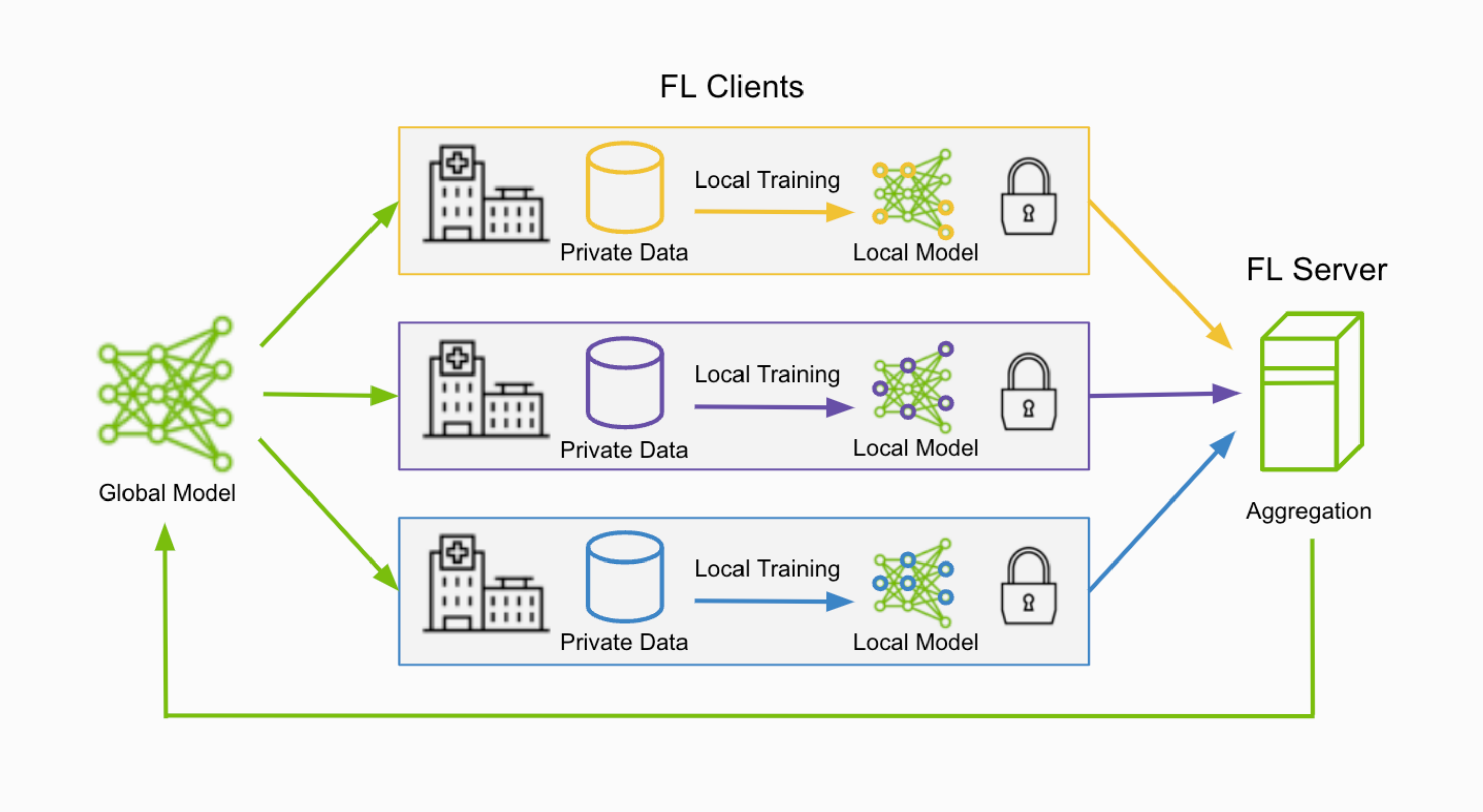}
    \caption{A visual representation of the federated learning protocol \cite{b3}. Each round of federated learning starts with a global model with initialized weights. Then, the FL server sends the global model to the clients. Next, the clients train locally on silo or device specific data and submit local updates only to the FL server. Then, the FL server performs some aggregation method on those local models to update the global model. Finally, the FL server sends the aggregated model back to the clients.}
    \label{fig:fl_diagram}
\end{figure}

Given the macroeconomic dependency of certain financial applications such as loan approvals \cite{b34}, institutions can greatly benefit from this collective intelligence to make more informed decisions. But this kind of training comes with several challenges, especially when considering the rapid development of FL. As shown in Fig. \ref{fig:fl_diagram}, in a standard federated learning protocol, the server sends a global model to selected institutions as local models and later sends their trained model updates back to the server. After this, the server updates the global model by aggregating the local updates to later send back to the institutions.  

Throughout this process, the individual institution's data can only be accessed by that institution, inherently ensuring data privacy. However, recent research identified several security risks in federated learning, such as data reconstruction and backdoor attacks \cite{b4,b5}. Despite these challenges, there has been promising progress in leveraging federated learning to enhance financial security. For instance, studies have explored blockchain-integrated cross-silo frameworks as a means to strengthen privacy and accountability \cite{b6}. In the area of fraud detection, FL has shown the potential to significantly improve accuracy by utilizing the collective insights of multiple institutions' data without compromising individual user privacy \cite{b7}. Additional research has demonstrated the broader impact of federated learning in financial security, including detecting counterfeit data through blockchain-based frameworks \cite{b8}, addressing data heterogeneity across clients via federated transfer learning \cite{b9}, and integrating quantum computing with federated learning for advanced financial services \cite{b10}.

Given the evolution and interdisciplinary nature of this field, a comprehensive survey is needed to consolidate knowledge and discuss emerging trends and challenges. This survey aims to fill that gap by organizing and analyzing existing research on federated learning in financial security, with a novel focus on classifying methodologies according to their regulatory and compliance exposure levels:
\begin{itemize}
    \item \textbf{Low-exposure:} e.g., credit risk assessment
    \item \textbf{Moderate-exposure:} e.g., loan approvals
    \item \textbf{High-exposure:} e.g., real-time fraud detection
\end{itemize}

By structuring the survey this way, the goal is to offer clarity on the suitability, effectiveness, and limitations of various FL approaches based on how critical or sensitive the financial application can be. This classification not only helps researchers and practitioners navigate a wide range of FL solutions more effectively but also highlights areas that require further exploration. Ultimately, this survey contributes to research in FL by mapping out the current state of FL in financial security and proposing directions for future research.

The remainder of this paper is organized as follows. Section \ref{sec:BG} reviews prior surveys and establishes the background for our study. Section \ref{sec:classification} describes a proposed classification of regulatory and compliance exposure. Section \ref{sec:FL_challenges} presents real-world challenges that Federated Learning faces in finance. A brief discussion of the ethics and social implications of Federated Learning in finance is provided in section \ref{sec:ethics}. Finally, section \ref{sec:future} discusses possible future practices of Federated Learning in financial security, and section \ref{sec:conclusion} concludes the paper.

\section{Background and Fundamentals}
\label{sec:BG}
Machine Learning (ML) and Artificial Intelligence (AI) has consumed our everyday life now more than ever. Television has advertisements about AI, some businesses run on AI, and most prominently Large Language Models (LLMs) such as ChatGPT has sparked a revolution of using AI not only in research but in many industries and even everyday life.

Financial services, banking, and insurance remain one of the biggest sectors that has a very high potential in utilizing ML with the availability of rich data and innovative algorithms \cite{b11}. In fact, there has already been rapid development in the usage of this technology in things such as risk modeling and stock portfolio management to algorithmic trading and fraud detection. However, traditional ML approaches often rely on centralized data collection, which raises serious concerns around privacy, security, and regulatory compliance. Centralizing sensitive financial data increases the risk of breaches, limits data sharing between institutions due to legal restrictions, and complicates compliance with regulations like GDPR \cite{b17}. Moreover, the "black-box" nature of traditional ML models makes analysis of their outputs and internal operations hinder acceptance and deployment into the real-world \cite{b11}.

\subsection{Federated Learning}

Federated learning (FL) is an alternative to traditional ML that allows multiple institutions to collaboratively train a shared model without transferring their raw data. In this decentralized framework, each participant such as a bank or financial institution trains the model locally on its own dataset and generates model updates (such as gradients or weights). The server aggregates these updates from all participants using algorithms like Federated Averaging \cite{b14} to update the global model, which is then redistributed to the participants for further training in successive rounds. This process repeats until the global model converges.

To further enhance privacy and security, secure aggregation \cite{b12} and differential privacy \cite{b13} are often incorporated FL systems. Secure aggregation ensures that individual model updates remain confidential by using cryptographic techniques to combine updates in such a way that only the final aggregated result is visible. This prevents any single update from being traced back to a participant. Differential privacy, on the other hand, introduces Gaussian noise into the updates before they are shared. This technique provides formal privacy guarantees, making it mathematically improbable to infer sensitive information about any single data point from the shared model. Together, these defenses make FL a robust, privacy-preserving approach for sensitive industries like finance. This allows financial institutions to collaborate more effectively in areas like fraud detection, risk assessment, and loan approvals without violating data protection laws\cite{b2}.

\subsection{Prior Surveys}
Current surveys have addressed the progress, challenges, and threats within the field of Federated Learning. In fact, \cite{b1} focus exclusively on the security threats in FL, including backdoor, Byzantine, and adversarial attacks. This work is valuable in showing the vulnerabilities that come with decentralized learning systems and the countermeasures being developed. However, their scope remains large, and fails to explore the domain-specific considerations such as financial applications and the regulations and risks that can come with those applications.

Another comprehensive survey is provided by \cite{b35} who conduct a holistic review of FL challenges, design, and application domains. They classify prior work by core system challenges and emerging aggregation and personalization techniques. However, their survey heavily focuses on FL applications in fields such as healthcare, mobile devices, and IoT. In fact, exposure classification is entirely excluded from their survey which we show to be extremely relevant in Financial applications. 

None of the surveyed literature addresses how FL frameworks can be customized to meet the strict demands of finance regulations, privacy concerns, and exposure classifications, where models must be accurate, explainable, and even auditable. While some works address general statistical heterogeneity and personalization in FL, they do not correlate these challenges to the implications they pose in finance, such as bias in credit scoring models or adversarial manipulation of loan approval systems. 

This work aims to fill the gap in this overlooked domain of FL applications. This survey provides a focused investigation into how FL can be adapted for varying levels of financial exposure in financial applications. By doing so, researchers can have an outline for deploying federated learning in finance.

\section{Regulatory and Compliance Exposure Classification}
\label{sec:classification}
This section categorizes FL applications according to their regulatory and compliance exposure. Regulatory and compliance exposure is the institution's exposure to potential legal or regulatory sanctions, loss of reputation, and financial loss due to bad practice of topics like data protection or failure to comply with internal policies or laws and regulations \cite{b38}. Compliance risk refers to an event with the likelihood of potential regulatory, financial, or reputational losses for the institution due to noncompliance \cite{b38}. We discuss the different levels of regulatory and compliance exposure below, and present different FL applications under each level along with key aspects and associated challenges. A summary of these applications is provided at the end of this section in Table \ref{table:fl_apps_summary}.


\subsection{Low-Exposure}
Low-exposure applications are characterized by minimal immediate regulatory oversight and involve compliance obligations that are primarily internal in nature or based on voluntarily adopted codes of conduct \cite{b38}. These applications typically involve personal decision making support systems where model outputs guide long-term decision making such as credit risk assessment \cite{b15} or collaborative portfolio optimization \cite{b16}. Generally, oversight in internal policies or best practices may lead to reputational risk, but the probability and impact of legal sanctions are minimal \cite{b38}.

\medskip
\subsubsection{Credit Risk Assessment}
Credit risk assessment is an easy-to-understand example of a low-exposure yet impactful application of FL in finance. A recent study proposed a FL prototype system where multiple financial institutions could collaboratively predict mortgage default risk without sharing raw data\cite{b15}. Using the Freddie Mac Single-Family Loan-Level dataset and economic indicators like delinquency, charge-off, and interest rate levels in the US at the national level \cite{b15}, the authors trained 14 simulated financial institutions on a time-series LSTM model. Under the assumption of Federated Learning architecture and FedAvg, the study found that data locality and privacy was preserved, while also enabling smaller financial institutions (smaller datasets) benefited from shared insights. In their experimentation, five scenarios were tested including setups with and without major contributors. The empirical results of this study showed that the FL model consistently outperformed both local (individual financial institutions) and centralized models, achieving a 99.04\% accuracy and a 98.22\% F1 score, compared to 95.04\% accuracy and 92.65\% F1 score on local models, even with the largest data contributors excluded.

While these numbers demonstrate FL's potential, the near-perfect performance may raise questions about their validity in today's world. The experiment relied on historical mortgage data curated under specific conditions (2006-2009 mortgage crisis), which may not reflect today's more diverse credit environment. Although the exclusion of major data contributors did not affect performance, real-world FL networks may experience more volatility due to uneven data quality or client dropouts. However, the study shows promise but should be interpreted as an optimistic upper bound rather than a guaranteed result.

These findings demonstrate FL's robustness and scalability while preserving regulatory compliance. The study further emphasized that the model performance correlated strongly with data volume, underscoring the value of collaboration among smaller institutions. This work serves as an empirical demonstration of FL in credit risk modeling using real mortgage portfolios, highlighting FL’s potential to enhance risk estimation without compromising privacy.

\medskip
\subsubsection{Collaborative Portfolio Optimization}
Collaborative Portfolio Optimization is another interesting low-exposure application of Federated Learning in Finance, enabling financial institutions to improve investment strategies while maintaining data privacy. A study conducted in December 2024 explores the concept of applying FL to the portfolio optimization process, which traditionally relies on centralized models constrained by fragmented data and regulatory limitations \cite{b16}. The authors frame FL as a solution that allows multiple institutions to co-train on local models on local datasets, preserving the confidentiality of proprietary trading data, client profiles, and risk metrics. Much like the risk assessment study\cite{b15}, this study shows that collaboration is especially beneficial to institutions with limited data because it allows them to leverage the macro view of the market without compromising compliance with regulations like GDPR \cite{b17} or CCPA \cite{b18}.

This study designed a simulation framework where several institutions collaboratively train optimization models without sharing raw data. Each institution optimizes portfolios locally using techniques based on Modern Portfolio Theory (MPT) \cite{b19}, while the global model updates through iterative weight aggregation via FedAvg. The proposed architecture supported varying data formats by allowing flexible aggregation techniques. Importantly, the author addressed challenges like data heterogeneity and computational complexity by distributing the training workload and enabling real-time adaptability in portfolio updates. 

Unlike the risk assessment study\cite{b15}, this study did not provide experimentation metrics or benchmarks comparisons but contributes significantly by demonstrating a conceptual and technical framework for implementing FL in portfolio management. It highlights how FL can enhance model robustness, simulate diverse market conditions, and reduce systemic risk \cite{b16}. Compared to traditional ML approaches, the proposed FL model offers scalability, privacy, and resilience without sacrificing the quality of portfolio decisions; setting the stage for future empirical studies, encouraging the adoption of privacy-preserving, collaborative optimization in portfolio management.

\subsection{Moderate-Exposure}
Moderate-exposure applications involve regulatory obligations, often requiring compliance with laws, regulations, rules, and codes of conduct applicable to the application \cite{b38}. This kind of exposure typically involves sanctions, financial loss, and most importantly increased monitoring and corrective measures \cite{b38}. These applications typically analyze sensitive data like income, credit behavior, employment history, and debt profiles. Traditional ML approaches for such predictions raise concerns over data breaches, lack of transparency, and inflexibility toward dynamic customer profiles. Federated Learning (FL) offers a promising alternative by enabling decentralized training of predictive models across multiple financial institutions without requiring raw data sharing. 

A study published in December 2024 proposes and Federated Machine Learning and Explainable AI (FML-XAI) framework designed for dynamic customer loan predictions \cite{b20}. By utilizing SHAP \cite{b21} and LIME \cite{b22} the Explainable AI (XAI) portion of this framework aims to offer interpretable insights into key drivers of model decisions like credit score, debt-to-income ratio, and income in hopes of building customer trust. This study directly addresses several research gaps such as:
\begin{itemize}
    \item \textbf {Privacy and Transparency:} FML lacks interpretability. XAI lacks privacy preservation. 
    \item \textbf{Dynamic Data Handling:} Most studies rely on static data, typically struggling in real-world deployment.
    \item \textbf{Scalability and Efficiency:} Existing FML systems are not scalable or communication efficient.
    \item \textbf{Domain-Specific Needs:} Financial applications require regulatory compliance and customer trust, which are not addressed in FML or XAI.
\end{itemize}

Implementing their own framework to combat these issues \cite{b20}, the authors demonstrate a solution that addresses these issues. Adaptability is addressed with real-time data drift\cite{b36}, where customer profiles change continuously due to changes in income, employment, spending patterns, and credit behavior. By using incremental learning and online federated learning, the proposed framework enables local models to incorporate new data while the global model evolves to understand emerging trends so predictions remain relevant. In addition to its adaptability, the framework offers strong privacy guarantees by ensuring that sensitive customer data never leaves each financial institution using FML. Transparency is addressed by using XAI techniques to provide interpretable explanations for model predictions. In their experimentation, the authors demonstrate the framework's scalbility and communication efficiency. Using FedAvg, along with optional differential privacy techniques, their system supports deployment across diverse financial institutions that may have differing capabilities and customer bases. In their results, model accuracy improved from 84\% to 92\% across federated rounds, while communication costs remained manageable; proving that the framework generalizes well across non-IID datasets and scales efficiently for real-world deployment. In short, by combining FML and XAI together their proposed framework ultimately enhances privacy, interpretability, adaptability, and scalability, making it ideal for modern loan approval systems.

\subsection{High-Exposure}
High-exposure applications occur in heavily regulated domains and typically carry not only regulatory and compliance risk but also integrity risks which may need continuous monitoring of changes in the regulatory environment to prevent exposure to crippling legal sanctions, substantial financial loss, or even detrimental reputational loss \cite{b38}. These applications typically involve real-time fraud detection \cite{b7,b8}, securing financial transactions \cite{b23}, and even Anti-Money Laundering (AML) \cite{b23,b24}. Federated Learning has become a focus in this research as a way to preserve privacy while enabling collaborative intelligence.

\smallskip
\subsubsection{Anti-Money Laundering}
Anti-Money Laundering (AML) compliance refers to the legal and regulatory processes financial institutions follow to detect, prevent, and report suspicious activities that aim to disguise the origins of illegally obtained money \cite{b25}. AML frameworks typically involve monitoring large volumes of financial transactions for anomalous behavior. However, existing AML systems are often limited by their reliance on centralized models, which are limited by incomplete data and usually produce false positives; estimates suggest that over 98\% of alerts are false, leading to increased operational costs \cite{b24}. Much like other financial applications, AML also has privacy regulation concerns  which prevent financial institutions from sharing sensitive customer transaction data. 

\begin{figure}[htbp]
    \centering
    \includegraphics[width=\columnwidth]{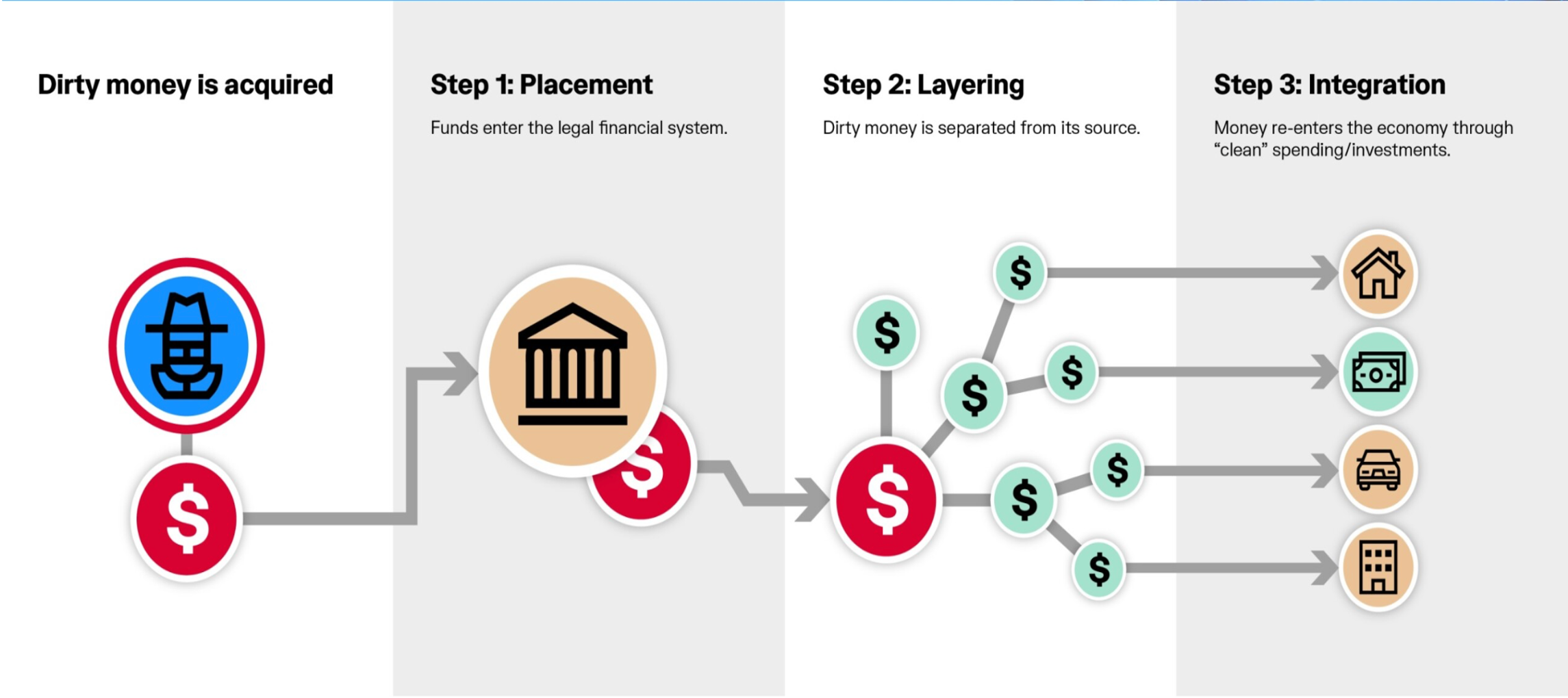}
    \caption{A visual representation of the money laundering schemes. Image credit: Linkurious, available at \url{https://linkurious.com/money-laundering-schemes/}}
    \label{fig:money_laundering_diagram}
\end{figure}

Federated Learning has come to light as a possible solution to these concerns by enabling multiple financial institutions to collaboratively train ML models on decentralized data, preserving data privacy while improving model performance through exposure to diverse patterns of money laundering. As we can see in Figure~\ref{fig:money_laundering_diagram}, a money launderer starts with dirty money that is simply entered into the legal financial system. However, in order to evade detection, the adversary may disperse these funds to several different sources. These sources could be foreign bank accounts, clean family members, shell companies, and even several domestic bank accounts. Once the money has been layered through sources, the adversary is able to actually integrate or spend their "clean" money back into the economy with no criminal backlash from having illegally obtained money. Intuitively, Federated Learning can be used to directly combat the layering method of money laundering by leveraging combined intelligence to make more informed decisions. For example, JP Morgan and Chase may partner with other domestic banks as well as foreign banks like BNP Paribas to gain a wider range of knowledge in where the money is going.

A recent study \cite{b24} proposed a novel FL-based AML detection framework that addresses the shortcomings of traditional systems. Inspired by the Federated Artificial Intelligence Enabler (FATE) framework and SecureBoost\cite{b37 }, the authors introduce a system where local models are trained individually at five virtual deposit banks and are coordinated through a central server without exchanging data. Each bank trains its model on local data using filter-based feature selection approach using XGBoost to highlight relevant transaction characteristics \cite{b25}. The models are encrypted and enhanced with differential privacy to ensure secure model updates. The architecture is supported by a formal privacy risk quantification using $\epsilon$ (privacy loss) and $\delta$ (risk of accidental data exposure), as well as the FedAvg protocol that aggregates model weights. 

In their experimentation, the authors discuss results that show the practicality and efficiency of their approach. The PaySim dataset was used, which contained 6 million mobile transactions, including both real and fraudulent behavior. The centralized baseline model achieved an accuracy of 0.99\% with a false alarm rate of 0.1\%. The classification report further indicates near-perfect scores, with an F1-score of 1.0 for benign transactions and 0.99 for malicious ones. The federated model produced results similar to the baseline model, occasionally lowering performance by a small fraction but still achieving an F1 score close to 1.0. While these findings suggest that FL can match centralized performance while preserving privacy, they should be interpreted with caution. The PaySim dataset is synthetic and highly controlled, which could bring on more desirable results compared to noisy, imbalanced, or even adversarial datasets encountered in real-world AML. Also, the marginal improvements of the federated model over the centralized baseline model suggest that FL's value lies less in accuracy gains and more in enabling privacy-preserving collaboration across institutions. Regulators may also be skeptical of near-perfect performance metrics, which often are a result of overfitting.

Although the FL approach seems to have no utility in terms of significant accuracy improvements of false positive rates, it is important to realize the actual importance of this study. These findings confirm that FL can maintain detection accuracy while overcoming concerns of data privacy, false positives, and model scalability. Ultimately, this approach shows FL's potential in the future of AML compliance, reducing operational costs and improving collaboration without compromising sensitive data. 

\begin{table*}[ht]
\centering
\caption{Summary of Federated Learning Applications in Finance}
\begin{tabular}{|p{2cm}|p{1cm}|p{4cm}|p{3.5cm}|p{4.5cm}|}
\hline
\textbf{Application} & \textbf{Exposure Level} & \textbf{Reported Performance} & \textbf{Privacy Techniques} & \textbf{Key Limitations} \\
\hline
Credit Risk Assessment & Low & 99.04\% accuracy, 98.22 F1 vs. 95.04\% / 92.65\% local models & FedAvg & Results tied to mortgage crisis data (2006–2009)\\
\hline
Collaborative Portfolio Optimization & Low & Conceptual framework with FedAvg aggregation across institutions &FedAvg + Secure Aggregation & No quantitative benchmarks; mainly theoretical \\
\hline
Loan Approvals & Moderate & Accuracy improved from 84\% to 92\% across federated rounds & FedAvg + Differential Privacy + Explainable AI (SHAP/LIME) & Validation limited to simulation; heterogeneous data challenges \\
\hline
Secure Financial Transactions & High & Privacy-preserving verification of transactions & FedAvg + Encryption & Performance not fully explored \\
\hline
Anti-Money Laundering & High & Centralized performance equal to federated performance but promise of privacy preserving and accuracy preserving collaboration & FATE + SecureBoost + Differential Privacy & Synthetic dataset (PaySim); “perfect” scores unlikely in practice \\
\hline
Fraud Detection using Blockchain & High & Detection accuracy improved with FL & Differential Privacy & High false positives inflate compliance costs \\
\hline
\end{tabular}
\label{table:fl_apps_summary}
\end{table*}

\medskip
\subsubsection{Fraud Detection with Blockchain-Based FL}
\label{sec:fraud_detection_blockchain}
Blockchain is a decentralized and distributed ledger system that records transactions across multiple nodes in a tamper-proof and transparent manner \cite{b26}. Each transaction is grouped into a block that is cryptographically secured, and linked to the previous one which makes a permanent chain. Once data is written onto the blockchain, it cannot be altered without consensus from the entire network. In the context of financial security, this inflexibility ensures that all interactions such as exchanges, model updates, or access permissions are verifiable and permanently recorded.

Traditional fraud detection models often rely on centralized data aggregation, which poses privacy risks and regulatory challenges. A comprehensive and innovative framework that utilizes FL with blockchain technology to enhance the security and, more importantly, the integrity of financial institutions is presented in \cite{b8}. The proposed framework uses blockchain's trustless environment and smart contract capabilities to improve the security and regulation of the FL process. Smart contracts are programmable rules deployed on the blockchain that automatically enforce access control, authentication, and model-sharing policies \cite{b8,b26}. This removes the need for intermediaries or centralized authorities, reducing the risks of manipulation, data leakage, or unauthorized access. The blockchain layer ensures that every institution in the FL system follows predefined protocols via smart contracts to ensure secure and accountable collaboration.

This blockchain-based FL architecture is designed to detect counterfeit financial transactions, including credit card fraud and application fraud. Smart contracts authenticate clients, authorize model submissions, manage version control, and securely synchronize local and global models \cite{b8}. This system guarantees that each update in the federated network is traceable, tamper-resistant, and compliant. The authors validate their proposed system using a synthetic dataset from Kaggle, which simulates 1.8 million real-world credit card transactions. The data includes 20 features, of which 15 were chosen. Six ML models were evaluated ( KNN, SVM, Decision Trees, Naive Bayes, Logistic Regression, and Random Forest), and the framework demonstrated promising performance in identifying fraudulent behavior while maintaining privacy standards. By combining blockchain's integrity with FL's privacy-preserving model training, the framework allows institutions to share knowledge without sharing sensitive data.


\section{Challenges of FL in Finance}
\label{sec:FL_challenges}
As discussed throughout this survey, FL offers compelling advantages for the financial sector, particularly in preserving private data while enabling collaborative intelligence. However, its adoption is not without hurdles. Federated systems must consider unique constraints and risks that are magnified in financial contexts due to high risks in decision outcomes and sensitive data environments.

This section categorizes the key challenges facing FL in finance into three areas: (1) technical limitations related to data heterogeneity and communication, (2) vulnerability to security threats despite privacy enhancements, and (3) regulatory and trust-related concerns that impact deployment and adoption.

\subsection{Data Heterogenity and Communication Overhead}
One of the biggest obstacles FL systems face is data heterogeneity and non-IID distributions. Financial institutions serve different demographics, operate in different regions, and handle diverse transaction types, leading to heterogeneous data distributions. This undermines global model convergence and reduces generalization performance, making it difficult to produce robust models that perform uniformly across all clients \cite{b27}. 

Communication overhead and scalability pose additional concerns. FL relies on iterative exchanges of model parameters between clients and a central server. In cross-silo settings, where financial institutions may have different infrastructure or bandwidth capabilities, communication can become resource-intensive and may introduce latency, limiting FL's suitability to time-sensitive financial tasks mentioned in this survey.

\subsection{Security Threats and Privacy Vulnerabilities}
Most importantly, although FL improves privacy by design, it remains vulnerable to security threats such as model poisoning \cite{b1} and backdoor attacks \cite{b5}. In model poisoning, malicious participants create and submit manipulated model updates that, when aggregated, degrade the overall performance or inject hidden objectives into the global model. A particular form is the backdoor attack, where the adversary trains their local model on poisoned data or alters weights such that the global model behaves normally on typical inputs but produces attacker-specified outputs when triggered by specific patterns \cite{b5} or inputs (misclassify transactions involving a certain account number). These attacks can be difficult to detect due to secure aggregation which masks individual updates to preserve privacy but also hides malicious behavior unintentionally. In the context of financial security, attackers could manipulate a fraud detection model to consistently ignore activities like money laundering associated with certain triggers like their account number. This allows the attacker to exploit things like fraud detection and money laundering while remaining hidden.

\subsection{Regulation Roadblocks and Trust}
Also, regulatory compliance is also an issue. Techniques like differential privacy offer formal privacy guarantees, but aligning these with financial industry regulations can become very tedious and complex. Transparency and consent is often demanded by regulators and current FL systems may not fully support these demands. 

Lastly, interpretability and trust are essential in financial applications where model decisions affect financial decisions, loan approvals, and fraud detection. Allowing AI models to make such vital decisions without any explanation of how they got the answer has hindered FL deployment in finance and could even ruin user trust.

\section{Ethical and Social Implications of FL in Finance}
\label{sec:ethics}
While the technical performance of FL frameworks is promising, its deployment in finance also raises broader ethical and societal considerations that extend beyond accuracy metrics.

\textbf{Transparency and Interpretability:} Current financial regulations often require that risk models be explainable to auditors, regulators, and more importantly, end users. FL frameworks that rely on deep learning may improve accuracy but risk reproducing the "black box" problem. Lack of interpretability may undermine customer trust, particularly in applications like credit scoring or loan approvals, where individual livelihoods are at risk.

\textbf{Fairness and Bias:} Credit risk datasets often reflect historical bias, such as regional discrepancies in mortgage approvals. FL may amplify these biases if the participating institutions contribute uneven data. Without fairness audits, the benefits of FL could unintentionally reinforce discrimination, creating ethical risks.

\textbf{Accountability:} Like any machine learning techniques, FL systems will make mistakes, and the question of responsibility can become complex. If multiple banks collaborate on a shared model, accountability for false positives in anti-money laundering or poor credit rejections could be distributed across all of the institutions and even the FL platform. This distribution of responsibility creates not only ethical challenges but also legal challenges.

Addressing these challenges requires not only technical solutions, such as Explainable AI and fairness-aware training, but also new regulations that could redefine accountability in collaborative systems.

\section{The Future of FL in Financial Security}
\label{sec:future}
Addressing these challenges requires a evolution of FL frameworks tailored to the needs of the financial sector.

\subsection{Blockchain-FL Integration}
A particularly promising direction is the integration of FL with blockchain technology and smart contracts, as explored in Section~\ref{sec:fraud_detection_blockchain}. Blockchain’s immutable ledger provides a secure and interpretable record of model updates, enhancing auditability and trust among stakeholders. Otherwise, smart contracts can be employed to automatically enforce compliance and participation rules, reducing the need for manual oversight and minimizing the potential for human error or adversarial attacks.

\subsection{Quantum Computing \& FL}
With the development and growth of quantum computing, traditional cryptographic methods used in current FL systems could become obsolete in years or decades to come. As such, developing quantum secure FL frameworks becomes essential. Quantum computers utilize things such as superposition and entanglement \cite{b28} to perform complex computations exponentially faster than classical computers. This capability opens doors for advanced financial analytics simulations for risk analysis and rapid portfolio optimization. However, it also threaten to break traditional encryption protocols such as RSA and elliptic curve cryptography which are foundational to the security of FL communications and data privacy \cite{b10}.

Beyond security, quantum computing could also enhance the computational efficiency of FL itself. Quantum machine learning algorithms could be adapted to federated settings, enabling clients to perform faster local training or optimization tasks which may reduce convergence time and communication rounds. While still an area of active research, hybrid classical-quantum FL architectures \cite{b10} could become useful in the future, especially in high-frequency trading where speed and accuracy are important. However, these advancements require significant infrastructural and practical developments in quantum computing that are simply many years away from achieving. In 2025, several tech companies such as Google\cite{b39}, IBM\cite{b40}, and Microsoft\cite{b41} have made large advancements in the Quantum Computing realm, but financial institutions are nowhere close to being quantum-ready. Ultimately, Quantum Computing in FL is a great way to combine the speed of Quantum Computing with the privacy guarantees of FL, but will require significant advancements in both fields of research.

\section{Conclusion}
\label{sec:conclusion}
Federated Learning represents a shift in how financial institutions can collaborate on machine learning models while preserving data privacy and regulatory compliance. By enabling decentralized model training, FL minimizes risks associated with centralized data and creates opportunities for secure, scalable AI systems in finance. The analysis across varying exposure levels (credit scoring to fraud detection) demonstrates FL’s broad use-cases and potential for real-world impact. However, issues like data heterogeneity, adversarial threats, and the lack of interpretability continue to hinder deployment. Also, the need for integration with regulations and the demand for trust and transparency from users must be addressed.

In particular, the rise of IoT-enabled financial endpoints such as POS systems and mobile banking devices make privacy-preserving machine learning solutions a necessity. Cross-device FL offers a compelling solution to address these needs, allowing models to learn directly from the endpoints without exposing sensitive data. Looking forward, the fusion of FL with blockchain technology presents an exciting opportunity for verifiable collaboration among financial institutions. Additionally, the rise of Quantum Computing poses both a threat and an opportunity. Quantum-secure FL architectures offer the promise of exponentially better computation speed. As these technologies evolve, the future of FL in finance will depend on efforts across machine learning, cryptography, financial regulation, and distributed learning. With the right frameworks and safeguards in place, FL can become a cornerstone of secure, intelligent, and ethical financial innovation.

\end{document}